\documentclass[10pt]{cernrep}
\usepackage{graphicx}
\usepackage{floatflt}

\newcommand{\Pt}{{P_t}}
\newcommand{\dphi}{\Delta\phi}
\newcommand{\phigj}{\phi_{(\gamma,jet)}}

\newcommand{\gpj}{~"$\gamma+Jet$"~}

\newcommand{\Db}{\Pt(O\!\!+\!\!\eta>5)}
\newcommand{\ptgj}{$~\Pt^{\gamma}$ and $\Pt^{Jet}~$}

\newcommand{\Ptg}{\Pt^{\gamma}}

\newcommand{\Ptgp}{Pt^{\gamma+part}}
\newcommand{\Ptgj}{Pt^{\gamma+jet}}
\newcommand{\Ptgje}{Pt^{\gamma+Jet}}
\newcommand{\Dpj}{Pt(O\!\!+\!\!\eta \! \!  > \! \! 5)/Pt^{\gamma}}

\newcommand{\gJ}{{(Pt^{\gamma}\! \! -\! \! Pt^{Jet})/Pt^{\gamma}}}

\suppressfloats[!]

\begin{document}


\thispagestyle{empty}
\thispagestyle{empty}

\vskip-5mm

\begin{center}
{\Large JOINT INSTITUTE FOR NUCLEAR RECEARCH}
\end{center}

\vskip10mm

\begin{flushright}
JINR Preprint \\
E2-2000-253 \\
hep-ex/0011084
\end{flushright}

\vspace*{3cm}

\begin{center}
\noindent
{\Large{\bfseries Jet energy scale setting with \gpj events at LHC \\[0pt]
energies. Minijets and cluster suppression and $\Pt^{\gamma}-\Pt^{Jet}$\\[0pt]
 disbalance.}}\\[5mm]
{\large D.V.~Bandourin$^{1\,\dag}$, V.F.~Konoplyanikov$^{2\,\ast}$, 
N.B.~Skachkov$^{3\,\dag}$}

\vskip 0mm

{\small
{\it
E-mail: (1) dmv@cv.jinr.ru, (2) kon@cv.jinr.ru, (3) skachkov@cv.jinr.ru}}\\[3mm]
$\dag$ \large \it Laboratory of Nuclear Problems \\
\hspace*{-4mm} $\ast$ \large \it Laboratory of Particle Physics
\end{center}

\vskip 9mm
\begin{center}
\begin{minipage}{150mm}
\centerline{\bf Abstract}
\noindent
In this paper the study of \gpj events is continued and the event number
determination useful for jet energy scale setting with the possible
 CMS detector hadron calorimeter calibration at LHC is fulfiled.
The sources of \ptgj are exposed. It is shown that the cluster $\Pt$
limitation, which is found in an event outside of the jet,
can be a good criterion for this disbalance decreasing.   
\end{minipage}
\end{center}

\newpage
\setcounter{page}{1}
\section{INTRODUCTION} 

In this article we continue our study of the $pp\to \gamma + Jet + X$~ process
caused by two partonic subprocesses:\\[-5mm]
\begin{eqnarray}
\hspace*{4.9cm} qg\to q+\gamma \hspace*{5.8cm} (1a)
\nonumber
\end{eqnarray}
\vspace{-7mm}
\begin{eqnarray}
\hspace*{4.9cm} q\overline{q}\to g+\gamma  \hspace*{5.8cm} (1b)
\nonumber
\end{eqnarray}
\setcounter{equation}{1}
\vspace{-4mm}

The main goal of this paper is to estimate whether
there  will be a sufficient number of \gpj events for setting
the mass scale of a jet with a good accuracy and for performing
hadron calorimeter (HCAL) calibration at LHC energies.
We use the PYTHIA generator as a model for this sort of estimation
supposing that the results with other physical event generators like
HERWIG, and with
GEANT--based simulation packages will be discussed in our further
 publications.

Here we study in detail the sources of \ptgj disbalance and the impact of
the $\Pt^{clust}_{CUT}$ parameter [1, 2], imposed as the cut on possible minijets or clusters
$\Pt$ for the calibration accuracy improvement
\footnote{
The detailed information on the dependence of the \ptgj disbalance on
$\Pt^{clust}_{CUT}$ and $\Pt^{out}_{CUT}$ in the case of background
events can be found in [3, 4].
}.
\section{DETAILED \gpj SYSTEM $\Pt$ DISBALANCE DEPENDENCE ON $\Pt^{clust}_{CUT}$ PARAMETER.}

In the previous papers ([1, 2]) we introduced physical observables
(variables) for studying \gpj events [1] and discussed what cuts for them
could lead to a decrease in the
$\Pt^{\gamma}$ and $\Pt^{Jet}$ disbalance [2].
The calibration  procedure depends on the gained statistics.
One can make these cuts to be tighter if more events would be collected
during data taking.

In the tables of Appendices 1--4 for four different $\Pt^{\gamma}$
intervals the mean values of the most important variables
that reflect the main features of \gpj events and
the other values that characterize \ptgj balance
(as predicted by the PYTHIA model), are presented.

Appendix 1 contains tables
for $\Pt^{\gamma}$ varying from 40 to
$50~GeV/c$. In these tables we present the values of
interest for three different Selections, mentioned in Section 3.2 [1].
Each
page corresponds to a definite value of $\Delta \phi$ (which enters 
the formula (23) of [1]) as a measure of deviation from the absolute
back-to-back orientation of two
$\vec{\Pt}^{\gamma}$ and $\vec{\Pt}^{Jet}$ vectors.
So, Tables 1 and 2 on the first page and Tables 3 and 4 on the second
one in each of Appendices 1--4 correspond to
$\dphi=180^{\circ}$ (i.e. to the case of no restriction
on the back-to-backness angle choice) and to $\dphi = 15^{\circ}$,
respectively. On the third page of each of Appendices 1--4 we present the Tables 5 and 6
(for the same cut $\dphi = 15^{\circ}$)
that correspond to the Selection 2 case described in [1].
This Selection differs from Selection 1, presented in Tables 1--4,
 by addition of the cut (26) of [1].
It allows one to select events with the ''isolated jet'', i.e. events
having the total $\Pt$ activity
in $\Delta R = 0.3$ ring around a jet not exceeding $2\%$
of jet $\Pt$. Tables 7, 8, shown on the fourth page in each of Appendices,
present results found in the case of Selection 3 for the same cut $\dphi=15^{\circ}$.
The last sample includes only events having the same jet found
simultaneously by both jetfinders UA1 and LUCELL.

The columns in Tables 1--6 correspond to five different
values of $Pt^{clust}_{CUT}=30,\ 20,\ 15,$ $10$ and $5 ~GeV/c$.
The upper lines of Tables 1--6 from all the Appendices
contain the numbers of events
$N_{event}$ expected for ``HB events'', i.e. for \gpj events in which jet is completely
fitted into the  barrel region of HCAL (see [2])
for the integrated luminosity
$L_{int}=3\;fb^{-1}$ for different values of $\Pt^{clust}_{CUT}$
and a fixed value of $\Delta \phi$. The last lines of the Tables
present the number of generated events, left after cuts, i.e. entries.
In the next four lines of the Tables we put the values of $\Pt56$,
$\Delta \phi$, $\Pt^{out}$, $\Pt^{\eta > 5}$
defined by formulae (3), (23), (25) and (5) of [1], respectively, and
averaged over the events selected with a chosen $\Pt^{clust}_{CUT}$ value.

From the Tables we see that the values of $\Pt56$,
$\Delta \phi$, $\Pt^{out}$ decrease fast
with decreasing $\Pt^{clust}_{CUT}\,$, while the averaged values of
$\Pt^{\eta>5}$ show very weak dependence on it (practically constant).

In the sixth line the average values of the
initial state vector disbalance $\Pt^{5+6}$ (defined by (3)
of [1]) are presented in addition to the rough scalar $\Pt56$ estimator value.
It is seen that they are smaller than $\Pt56$ value.

The next lines represent the average
values of the variables
$\Pt^{\gamma+part}$,
$\Pt^{\gamma+jet}$,
$\Pt^{\gamma+Jet}$,
that are defined as averaged values of (2), (12) and (4) from [1] and
serve as a measure of the $\Pt$ disbalance in the \gpj system.
These lines correspond in the following order to: the balance at the parton level,
the balance of that
part of the \gpj system which can be measured by calorimeters,
 i.e. defined by
$\Pt^{jet}$ (see (10) of [1]), and the balance of the ``$\gamma$ $+$
complete jet'' system. Practically all the values of these three variables
drop approximately by a factor of two when we move
from $\Pt^{clust}=30 ~GeV/c$ to $\Pt^{clust}=5 ~GeV/c$
for all $\Pt^{\gamma}$ intervals and
for both the UA1 and LUCELL algorithms.

To take into account the part of jet $\Pt$ carried off by neutrinos,
 we have introduced in [1]
the correction $\Delta_\nu$ that should be added to $\Pt^{jet}$ to restore
$\Pt$ of the complete jet in each event:
\begin{equation}
\Delta_\nu=\Pt^{Jet}-\Pt^{jet}
\label{eq:nu1}
\end{equation}
Let us define a new quantity $\Pt^J$, which would serve
as an estimator for the total $\Pt^{Jet}$ value, by the sum of $\Pt^{jet}$
(a measurable part of $\Pt^{Jet}$) and an averaged
$\Delta_\nu$ correction: \\[-2pt]
\begin{equation}
\Pt^J=\Pt^{jet}+<\!\!\Delta_\nu\!\!>
\label{eq:nu2}
\end{equation}

The comparison of the $\Pt^{\gamma+part}$ and $\Pt^{\gamma+Jet}$
shows that the fragmentation process contribution into the value
of the final state $\Pt$ disbalance is much more smaller than the
contribution of ISR that defines a dominant part of $\Pt$ disbalance
in the \gpj system. The photon and the jet $\Pt$ disbalance is defined in fact by
the disbalance appearing at the parton level of fundamental $2 \to 2$
subprocesses (1a) and (1b). The comparison of $\Pt^{\gamma+Jet}$ and
$\Pt^{5+6}$ shows that the final state disbalance has, approximately,
the value of the initial state $\Pt$ disbalance of colliding partons.

After the described lines follow three lines below  for the averaged values of the
$\,(\Pt^{\gamma}-\Pt^J)/\Pt^{\gamma}$, $\,(1-cos(\dphi))$ and $\Db/\Pt^{\gamma}$
quantities that enter 
equation (29) of [1] which has the meaning of the scalar variant of vector
equation (16) of [1] for the total transverse momentum conservation in a physical event.
The first quantity characterizes  relative $\Pt$ disbalance in the
``$\gamma$ $+$ complete Jet'' system.
The  second and the third ones have the meaning of the averaged values of
two terms in the
right-hand part of balance equation (29) of paper [1].
The value of $\left<(1-cos(\dphi))\right>$ is smaller than $\left<\Db/\Pt^{\gamma}\right>$
 for the cut
$\dphi\leq15^\circ$ and tends to decrease more with
a growth of the energy. So, we conclude that the main source of the $\Pt$ disbalance
in \gpj system is defined by the term $\left<\Db/\Pt^{\gamma}\right>$.

We see from the Tables  that
more restrictive cuts on
the $\Pt^{clust}$ observable
lead to decreasing in the $\Pt56$ and $\Pt^{5+6}$ variables
(non-observable ones)
that serve, according to (3) of paper [1], as
measures of the initial state radiation transverse momentum $\Pt^{ISR}$,
i.e. of the main source of the $\Pt$ disbalance in
fundamental $2\to 2$ subprocesses (1a) and (1b).

Thus, the variation of
$\Pt^{clust}_{CUT}$ from $30 ~GeV/c$ to
$5 ~GeV/c$ for $\dphi \leq 15^\circ$
leads to suppression of the $\Pt56$ and $\Pt^{5+6}$ values
(or $\Pt^{ISR}$) approximately by $25\%$ for $40<\Pt^{\gamma}<50 ~GeV/c$
and by $\approx 40-45\%$ for $\Pt^{\gamma} \geq 100 ~GeV/c$.
This diminishing of $\Pt^{clust}_{CUT}$ value leads to
diminishing of $(\Pt^{\gamma}-\Pt^J)/\Pt^{\gamma}$
ratio, i.e. we improve the calibration accuracy. For instance, in the
case of $100<\Pt^{\gamma}<120 ~GeV/c$ the mean value of
$(\Pt^{\gamma}\!-\!\Pt^J)/\Pt^{\gamma}$ drops from $4\!\!-\!\!5\%$ to
$0.6\!\!-\!\!1\%$ (see Table 3 and 4 of Appendix 2) and in the
case of $200<\Pt^{\gamma}<240 ~GeV/c$ the mean value of this variable
drops from $2\%$ to less than $0.5\%$ (see Tables 3 and 4 of Appendix 3).

After requirement the jet to be isolated (see Tables 5, 8 of Appendix 1--4)
we observe, starting from $\Pt^{\gamma} = 100 ~GeV/c$, that the mean values of
$(\Pt^{\gamma}\!-\!\Pt^J)/\Pt^{\gamma}$ are contained inside the $1\%$ window
for any $\Pt^{clust}$ value. In the case of $40<\Pt^{\gamma}<50 ~GeV/c$ interval,
where we have enough events even after passing to Selections 2 and 3, we see that
$\Pt^{clust}_{CUT}$ works most effectively. Thus, $\Pt^{clust}_{CUT}=10 ~ GeV/c$
allows to reduce $(\Pt^{\gamma}\!-\!\Pt^J)/\Pt^{\gamma}$ value to be less than $1.5\%$
in the case of Selection 3 while a more strict cut $\Pt^{clust}_{CUT}=5 ~ GeV/c$
makes it less than $0.5\%$. Both cuts leave quite a sufficient number of events:
about 124 and 35 thousand, correspondingly (see Tables 7, 8 of Appendix 1).

In the following papers [3, 4] we shall show how these results can be improved
by imposing the cut on $\Pt^{out}$ as it enter the expression $\Db/\Pt^{\gamma}$,
which gives a dominant contribution to the right-hand side of $\Pt$-balance equation
(29) of [1] as we have mentioned above.

\section{SUMMARY}
The efficiency of the $\Pt^{clust}$ restriction for the initial state radiation
(ISR) suppression and the $\Pt$ balance improvement in the \gpj system was demonstrated.
For the case of $\dphi\leq 15^\circ$ such a strict limitation as
$\Pt^{clust}_{CUT}=5 ~GeV/c$ would allow to reduce $\Pt^{ISR}$ by $25-40\%$
(in dependence on $\Ptg$) and simultaneously to obtain the $1\%$
disbalance value.

It is also shown that the number of the events (at $L_{int}=3 fb^{-1}$),
collected by Selection 2 criteria are not small even at
$\Pt^{clust}_{CUT}=5-10 ~GeV/c$. These events have a topologically clean jet,
whose transverse momentum is good balanced with one of the direct photon.

\section{ ACKNOWLEDGMENTS}                                         

We are greatly thankful to D.~Denegri for having offered this theme to study,
fruitful discussions and permanent support and encouragement.
It is a pleasure for us
to express our recognition for helpful discussions to P.~Aurenche,
M.~Dittmar, M.~Fontannaz, J.Ph.~Guillet, M.L.~Mangano, E.~Pilon,
H.~Rohringer, S.~Tapprogge and J.~Womersley.


\setcounter{table}{0}

\begin{table}
\noindent
\large{\bf Appendix 1}\\
\begin{center}
\large{ $40 < \Pt^{\gamma} < 50 ~GeV/c$}\\[20pt]
\normalsize
\caption{Selection 1. $ \phigj=180^\circ \pm  180^\circ$. UA1 algorithm. }
\vskip0.2cm
\begin{tabular}{||c||c|c|c|c|c|c|c|c|c|c||}
\hline
\hline
$Pt^{clust}_{CUT}$ &$\quad$  30 $\quad$&$\quad$  20 $\quad$&$\quad$ 15 $\quad$&$\quad$  10 $\quad$&$\quad\ $   5 $\quad\ $\\\hline
\hline
Nevent$^\ast$             &  2829803&  2319650&  1904754&  1283404&   296186   \\\hline
$Pt56$                    &     16.8&     14.0&     12.1&     10.0&      7.6   \\\hline
$\Delta \phi$             &     13.6&     10.3&      8.6&      6.8&      5.0   \\\hline
$Pt^{out}$                &     13.6&     11.1&      9.7&      8.0&      5.8   \\\hline
$Pt^{\eta > 5}$           &      4.6&      4.6&      4.5&      4.5&      4.3   \\\hline
$Pt^{5+6}$                &     14.0&     11.4&      9.7&      7.9&      6.0   \\\hline
$\Ptgp$&     13.9&     11.3&      9.7&      7.9&      6.0\\\hline
$\Ptgj$ &     13.3&     10.8&      9.4&      7.8&      6.0\\\hline
$\Ptgje$   &     13.2&     10.8&      9.3&      7.8&      6.0\\\hline
$\gJ$                     &   0.0410&   0.0432&   0.0413&   0.0351&   0.0220  \\\hline
$1-cos(\Delta \phi)$      &   0.0546&   0.0309&   0.0209&   0.0133&   0.0083   \\\hline
$\Dpj$                    &  -0.0135&   0.0124&   0.0205&   0.0218&   0.0138  \\\hline
                   Entries&    83981&    68841&    56528&    38088&     8790   \\\hline
\hline
\end{tabular}
\vskip0.8cm
\caption{Selection 1. $ \phigj=180^\circ \pm  180^\circ$. LUCELL algorithm. }
\vskip0.2cm
\begin{tabular}{||c||c|c|c|c|c|c|c|c|c|c||}
\hline
\hline
$Pt^{clust}_{CUT}$ &$\quad$  30 $\quad$&$\quad$  20 $\quad$&$\quad$ 15 $\quad$&$\quad$  10 $\quad$&$\quad\ $   5 $\quad\ $\\\hline
\hline
Nevent                    &  2739600&  2208386&  1781966&  1145083&   267241\\\hline
$Pt56$                    &     16.7&     13.7&     11.7&      9.4&      7.0\\\hline
$\Delta \phi$             &     13.4&     10.0&      8.1&      6.2&      4.5\\\hline
$Pt^{out}$                &     13.4&     10.8&      9.3&      7.5&      5.3\\\hline
$Pt^{\eta > 5}$           &      4.6&      4.6&      4.5&      4.4&      4.2\\\hline
$Pt^{5+6}$                &     13.9&     11.2&      9.4&      7.5&      5.4\\\hline
$\Ptgp$&     13.8&     11.1&      9.3&      7.4&      5.4\\\hline
$\Ptgj$ &     13.1&     10.5&      9.0&      7.2&      5.4\\\hline
$\Ptgje$   &     13.1&     10.5&      8.9&      7.2&      5.4\\\hline
$\gJ$                     &   0.0365&   0.0380&   0.0348&   0.0282&   0.0169\\\hline
$1-cos(\Delta \phi)$      &   0.0525&   0.0281&   0.0182&   0.0106&   0.0056\\\hline
$\Dpj$                    &  -0.0160&   0.0100&   0.0166&   0.0177&   0.0113\\\hline
                   Entries&    81304&    65539&    52884&    33983&     7931\\\hline
\hline
\end{tabular}
\end{center}
~\\[1cm]
\underline{~~~~~~~~~~~~~~~~~~~~~~~~~~~}\\
\hspace*{0.5cm}\footnotesize{~$^\ast$Number of events (Nevent) is given in this and
in the following tables for integrated luminosity $L_{int}=3~fb^{-1}$ }
\end{table}

\begin{table}
\begin{center}
\vskip-1cm
\caption{Selection 1. $ \phigj=180^\circ \pm  15^\circ$. UA1 algorithm. }
\vskip0.2cm
\begin{tabular}{||c||c|c|c|c|c|c|c|c|c|c||}
\hline
\hline
$Pt^{clust}_{CUT}$ &$\quad$  30 $\quad$&$\quad$  20 $\quad$&$\quad$ 15 $\quad$&$\quad$  10 $\quad$&$\quad\ $   5 $\quad\ $\\\hline
\hline
Nevent                    &  1904888&  1771992&  1583397&  1160010&   283752\\\hline
$Pt56$                    &     12.2&     11.3&     10.4&      9.0&      7.0\\\hline
$\Delta \phi$             &      5.9&      5.8&      5.7&      5.2&      4.2\\\hline
$Pt^{out}$                &      9.0&      8.6&      8.1&      7.2&      5.5\\\hline
$Pt^{\eta > 5}$           &      4.5&      4.4&      4.4&      4.3&      4.1\\\hline
$Pt^{5+6}$                &      9.3&      8.7&      8.0&      6.9&      5.4\\\hline
$\Ptgp$&      9.3&      8.6&      7.9&      6.9&      5.4\\\hline
$\Ptgj$ &      8.3&      7.9&      7.5&      6.7&      5.4\\\hline
$\Ptgje$   &      8.3&      7.9&      7.5&      6.7&      5.4\\\hline
$\gJ$                     &   0.0293&   0.0342&   0.0356&   0.0324&   0.0213\\\hline
$1-cos(\Delta \phi)$      &   0.0080&   0.0077&   0.0073&   0.0063&   0.0044\\\hline
$\Dpj$                    &   0.0214&   0.0266&   0.0283&   0.0261&   0.0170\\\hline
                   Entries&    56532&    52588&    46991&    34426&     8421\\\hline
\hline
\end{tabular}
\vskip0.8cm
\caption{Selection 1. $ \phigj=180^\circ \pm  15^\circ$. LUCELL algorithm. }
\vskip0.2cm
\begin{tabular}{||c||c|c|c|c|c|c|c|c|c|c||}
\hline
\hline
$Pt^{clust}_{CUT}$ &$\quad$  30 $\quad$&$\quad$  20 $\quad$&$\quad$ 15 $\quad$&$\quad$  10 $\quad$&$\quad\ $   5 $\quad\ $\\\hline
\hline
Nevent                    &  1850638&  1709150&  1507481&  1059900&   260259\\\hline
$Pt56$                    &     12.2&     11.2&     10.3&      8.8&      6.7\\\hline
$\Delta \phi$             &      6.0&      5.8&      5.6&      5.1&      4.1\\\hline
$Pt^{out}$                &      8.9&      8.4&      7.9&      7.0&      5.3\\\hline
$Pt^{\eta > 5}$           &      4.5&      4.4&      4.4&      4.3&      4.0\\\hline
$Pt^{5+6}$                &      9.4&      8.7&      7.9&      6.8&      5.2\\\hline
$\Ptgp$&      9.3&      8.6&      7.9&      6.8&      5.2\\\hline
$\Ptgj$ &      8.3&      7.8&      7.4&      6.5&      5.2\\\hline
$\Ptgje$   &      8.2&      7.8&      7.3&      6.5&      5.1\\\hline
$\gJ$                     &   0.0236&   0.0287&   0.0290&   0.0257&   0.0163\\\hline
$1-cos(\Delta \phi)$      &   0.0080&   0.0077&   0.0073&   0.0062&   0.0042\\\hline
$\Dpj$                    &   0.0156&   0.0210&   0.0218&   0.0196&   0.0122\\\hline
                   Entries&    54922&    50723&    44738&    31455&     7751\\\hline
\hline
\end{tabular}

\end{center}
\end{table}

\begin{table}
\begin{center}
\vskip-1cm
\caption{Selection 2. $ \phigj=180^\circ \pm  15^\circ$. UA1 algorithm. }
\vskip0.2cm
\begin{tabular}{||c||c|c|c|c|c|c|c|c|c|c||}
\hline
\hline
$Pt^{clust}_{CUT}$ &$\quad$  30 $\quad$&$\quad$  20 $\quad$&$\quad$ 15 $\quad$&$\quad$  10 $\quad$&$\quad\ $   5 $\quad\ $
\\\hline
\hline
Nevent                    &   222459&   208947&   190853&   150148&    52363\\\hline
$Pt56$                    &     11.4&     10.4&      9.6&      8.4&      6.7\\\hline
$\Delta \phi$             &      5.6&      5.5&      5.4&      4.9&      4.1\\\hline
$Pt^{out}$                &      8.9&      8.2&      7.7&      6.8&      5.3\\\hline
$Pt^{\eta > 5}$           &      4.5&      4.4&      4.4&      4.4&      4.2\\\hline
$Pt^{5+6}$                &      8.6&      7.9&      7.3&      6.4&      5.1\\\hline
$\Ptgp$&      8.7&      8.0&      7.4&      6.5&      5.2\\\hline
$\Ptgj$ &      8.3&      7.6&      7.1&      6.4&      5.3\\\hline
$\Ptgje$   &      8.3&      7.7&      7.2&      6.4&      5.3\\\hline
$\gJ$                     &  -0.0469&  -0.0336&  -0.0252&  -0.0187&  -0.0137\\\hline
$1-cos(\Delta \phi)$      &   0.0073&   0.0070&   0.0067&   0.0059&   0.0041\\\hline
$\Dpj$                    &  -0.0540&  -0.0405&  -0.0318&  -0.0245&  -0.0178\\\hline
                   Entries&     6602&     6201&     5664&     4456&     1554\\\hline
\hline
\end{tabular}
\vskip0.8cm
\caption{Selection 2. $ \phigj=180^\circ \pm  15^\circ$. LUCELL algorithm. }
\vskip0.2cm
\begin{tabular}{||c||c|c|c|c|c|c|c|c|c|c||}
\hline
\hline
$Pt^{clust}_{CUT}$ &$\quad$  30 $\quad$&$\quad$  20 $\quad$&$\quad$ 15 $\quad$&$\quad$  10 $\quad$&$\quad\ $   5 $\quad\ $\\\hline
\hline
Nevent                    &   219764&   205275&   185394&   140916&    46972\\\hline
$Pt56$                    &     11.3&     10.3&      9.4&      8.1&      6.4\\\hline
$\Delta \phi$             &      5.6&      5.5&      5.3&      4.9&      4.0\\\hline
$Pt^{out}$                &      8.9&      8.2&      7.6&      6.6&      5.1\\\hline
$Pt^{\eta > 5}$           &      4.5&      4.5&      4.4&      4.3&      4.1\\\hline
$Pt^{5+6}$                &      8.6&      7.8&      7.1&      6.2&      4.9\\\hline
$\Ptgp$&      8.7&      7.9&      7.2&      6.3&      4.9\\\hline
$\Ptgj$ &      8.2&      7.6&      7.1&      6.2&     5.0\\\hline
$\Ptgje$   &      8.3&      7.6&      7.1&      6.2&  5.0\\\hline
$\gJ$                     &  -0.0438&  -0.0289&  -0.0219&  -0.0148&  -0.0073\\\hline
$1-cos(\Delta \phi)$      &   0.0073&   0.0071&   0.0067&   0.0057&   0.0040\\\hline
$\Dpj$                    &  -0.0510&  -0.0358&  -0.0284&  -0.0204&  -0.0112\\\hline
                   Entries&     6522&     6092&     5502&     4182&     1398\\\hline
\hline
\end{tabular}
\end{center}
\end{table}

\begin{table}
\begin{center}
\vskip-1cm
\caption{Selection 3. $ \phigj=180^\circ \pm  15^\circ$. UA1 algorithm. }
\vskip0.2cm
\begin{tabular}{||c||c|c|c|c|c|c|c||}
\hline
\hline
$Pt^{clust}_{CUT}$ &$\quad$  30 $\quad$&$\quad$  20 $\quad$&$\quad$ 15 $\quad$&$\quad$  10 $\quad$&$\quad\ $   5 $\quad\ $\\\hline
\hline
Nevent                    &   204381&   190906&   170562&   124006&    35229\\\hline
$Pt56$                    &     10.8&      9.6&      8.7&      7.5&      5.9\\\hline
$\Delta \phi$             &      5.5&      5.3&      5.1&      4.6&      3.6\\\hline
$Pt^{out}$                &      8.5&      7.7&      7.1&      6.1&      4.3\\\hline
$Pt^{\eta > 5}$           &      4.5&      4.4&      4.3&      4.3&      4.0\\\hline
$Pt^{5+6}$                &      8.1&      7.2&      6.6&      5.7&      4.5\\\hline
$\Ptgp$&      8.3&      7.3&      6.7&      5.8&      4.5\\\hline
$\Ptgj$ &      7.9&      7.1&      6.6&      5.7&     4.5\\\hline
$\Ptgje$   &      7.9&      7.1&      6.6&      5.7&  4.5\\\hline
$\gJ$                     &  -0.0408&  -0.0268&  -0.0201&  -0.0142&  -0.0042\\\hline
$1-cos(\Delta \phi)$      &   0.0071&   0.0067&   0.0061&   0.0051&   0.0031\\\hline
$\Dpj$                    &  -0.0478&  -0.0334&  -0.0261&  -0.0192&  -0.0072\\\hline
                   Entries&     6065&     5604&     5061&     3680&     1049\\\hline
\hline
\end{tabular}
\vskip0.8cm
\caption{Selection 3. $ \phigj=180^\circ \pm  15^\circ$. LUCELL algorithm. }
\vskip0.2cm
\begin{tabular}{||c||c|c|c|c|c|c|c|c|c|c||}
\hline
\hline
$Pt^{clust}_{CUT}$ &$\quad$  30 $\quad$&$\quad$  20 $\quad$&$\quad$ 15 $\quad$&$\quad$  10 $\quad$&$\quad\ $   5 $\quad\ $\\\hline
\hline
Nevent                    &   204381&   190906&   170562&   124006&    35229\\\hline
$Pt56$                    &     10.8&      9.6&      8.7&      7.5&      5.9\\\hline
$\Delta \phi$             &      5.5&      5.3&      5.1&      4.6&      3.6\\\hline
$Pt^{out}$                &      8.5&      7.7&      7.1&      6.1&      4.3\\\hline
$Pt^{\eta > 5}$           &      4.5&      4.4&      4.3&      4.3&      4.0\\\hline
$Pt^{5+6}$                &      8.1&      7.2&      6.6&      5.7&      4.5\\\hline
$\Ptgp$&      8.3&      7.3&      6.7&      5.8&      4.5\\\hline
$\Ptgj$ &      7.9&      7.1&      6.6&      5.7&     4.6\\\hline
$\Ptgje$   &      7.9&      7.1&      6.6&      5.8&  4.6\\\hline
$\gJ$                     &  -0.0400&  -0.0260&  -0.0194&  -0.0136&  -0.0036\\\hline
$1-cos(\Delta \phi)$      &   0.0072&   0.0067&   0.0062&   0.0051&   0.0032\\\hline
$\Dpj$                    &  -0.0470&  -0.0326&  -0.0255&  -0.0186&  -0.0066\\\hline
                   Entries&     6065&     5604&     5061&     3680&     1049\\\hline
\hline
\end{tabular}
\end{center}
\end{table}

\setcounter{table}{0}
\begin{table}
\large{\bf Appendix 2}\\
\begin{center}
\large{ $100 < \Pt^{\gamma} < 120 ~GeV/c$}\\[20pt]
\normalsize
\caption{Selection 1. $ \phigj=180^\circ \pm  180^\circ$. UA1 algorithm. }
\vskip0.2cm
\begin{tabular}{||c||c|c|c|c|c|c|c|c|c||}
\hline
\hline
$Pt^{clust}_{CUT}$ &$\quad$  30 $\quad$&$\quad$  20 $\quad$&$\quad$ 15 $\quad$&$\quad$  10 $\quad$&$\quad\ $   5 $\quad\ $\\\hline
\hline
Nevent                    &   133709&    95415&    72927&    45345&     8654\\\hline
$Pt56$                    &     23.1&     17.7&     15.1&     12.6&      9.9\\\hline
$\Delta \phi$             &      6.4&      4.7&      3.9&      3.1&      2.4\\\hline
$Pt^{out}$                &     18.5&     13.6&     11.3&      9.1&      6.7\\\hline
$Pt^{\eta > 5}$           &      4.8&      4.8&      4.7&      4.7&      4.6\\\hline
$Pt^{5+6}$                &     19.3&     14.3&     12.1&     10.0&      7.9\\\hline
$\Ptgp$&     19.1&     14.3&     12.1&     10.1&      8.0\\\hline
$\Ptgj$ &     18.8&     14.0&     11.7&      9.6&      7.5\\\hline
$\Ptgje$   &     18.7&     13.9&     11.6&      9.5&      7.5\\\hline
$\gJ$                     &   0.0589&   0.0335&   0.0238&   0.0170&   0.0120\\\hline
$1-cos(\Delta \phi)$      &   0.0120&   0.0062&   0.0042&   0.0029&   0.0024\\\hline
$\Dpj$                    &   0.0470&   0.0273&   0.0196&   0.0141&   0.0097\\\hline
                   Entries&    69710&    49745&    38021&    23641&     4512\\\hline
\hline
\end{tabular}
\vskip0.8cm
\caption{Selection 1. $ \phigj=180^\circ \pm  180^\circ$. LUCELL algorithm. }
\vskip0.2cm
\begin{tabular}{||c||c|c|c|c|c|c|c|c|c|c||}
\hline
\hline
$Pt^{clust}_{CUT}$ &$\quad$  30 $\quad$&$\quad$  20 $\quad$&$\quad$ 15 $\quad$&$\quad$  10 $\quad$&$\quad\ $   5 $\quad\ $\\\hline
\hline
Nevent                    &   124587&    87645&    65928&    38555&     7720\\\hline
$Pt56$                    &     22.3&     16.9&     14.2&     11.4&      8.6\\\hline
$\Delta \phi$             &      6.1&      4.4&      3.6&      2.8&      2.0\\\hline
$Pt^{out}$                &     17.6&     12.8&     10.5&      8.1&      5.7\\\hline
$Pt^{\eta > 5}$           &      4.8&      4.7&      4.7&      4.7&      4.6\\\hline
$Pt^{5+6}$                &     18.5&     13.6&     11.3&      9.0&      6.7\\\hline
$\Ptgp$&     18.4&     13.6&     11.4&      9.1&      6.7\\\hline
$\Ptgj$ &     18.0&     13.2&     10.9&      8.5&      6.4\\\hline
$\Ptgje$   &     17.9&     13.1&     10.8&      8.5&      6.4\\\hline
$\gJ$                     &   0.0504&   0.0268&   0.0178&   0.0104&   0.0063\\\hline
$1-cos(\Delta \phi)$      &   0.0108&   0.0053&   0.0035&   0.0021&   0.0012\\\hline
$\Dpj$                    &   0.0397&   0.0215&   0.0143&   0.0083&   0.0051\\\hline
                   Entries&    64954&    45694&    34372&    20101&     4025\\\hline
\hline
\end{tabular}
\end{center}
\end{table}

\begin{table}
\begin{center}
\vskip-1cm
\caption{Selection 1. $ \phigj=180^\circ \pm  15^\circ$. UA1 algorithm. }
\vskip0.2cm
\begin{tabular}{||c||c|c|c|c|c|c|c|c|c|c||}
\hline
\hline
$Pt^{clust}_{CUT}$ &$\quad$  30 $\quad$&$\quad$  20 $\quad$&$\quad$ 15 $\quad$&$\quad$  10 $\quad$&$\quad\ $   5 $\quad\ $\\\hline
\hline
Nevent                    &   121445&    92409&    71951&    45021&     8568\\\hline
$Pt56$                    &     20.8&     16.9&     14.7&     12.3&      9.5\\\hline
$\Delta \phi$             &      4.9&      4.2&      3.7&      3.0&      2.2\\\hline
$Pt^{out}$                &     16.1&     12.9&     11.0&      8.9&      6.4\\\hline
$Pt^{\eta > 5}$           &      4.7&      4.7&      4.7&      4.7&      4.5\\\hline
$Pt^{5+6}$                &     16.9&     13.5&     11.7&      9.8&      7.5\\\hline
$\Ptgp$&     16.8&     13.6&     11.8&      9.8&      7.6\\\hline
$\Ptgj$ &     16.3&     13.1&     11.3&      9.3&      7.2\\\hline
$\Ptgje$   &     16.2&     13.1&     11.2&      9.3&      7.1\\\hline
$\gJ$                     &   0.0503&   0.0312&   0.0228&   0.0162&   0.0106\\\hline
$1-cos(\Delta \phi)$      &   0.0060&   0.0045&   0.0035&   0.0024&   0.0015\\\hline
$\Dpj$                    &   0.0443&   0.0268&   0.0194&   0.0138&   0.0092\\\hline
                   Entries&    63316&    48178&    37512&    23472&     4467\\\hline
\hline
\end{tabular}
\vskip0.8cm
\caption{Selection 1. $ \phigj=180^\circ \pm  15^\circ$. LUCELL algorithm. }
\vskip0.2cm
\begin{tabular}{||c||c|c|c|c|c|c|c|c|c|c||}
\hline
\hline
$Pt^{clust}_{CUT}$ &$\quad$  30 $\quad$&$\quad$  20 $\quad$&$\quad$ 15 $\quad$&$\quad$  10 $\quad$&$\quad\ $   5 $\quad\ $\\\hline
\hline
Nevent                    &   114477&    85721&    65481&    38500&     7703\\\hline
$Pt56$                    &     20.4&     16.4&     14.1&     11.4&      8.5\\\hline
$\Delta \phi$             &      4.9&      4.1&      3.5&      2.8&      2.0\\\hline
$Pt^{out}$                &     15.6&     12.3&     10.4&      8.1&      5.7\\\hline
$Pt^{\eta > 5}$           &      4.7&      4.7&      4.7&      4.6&      4.5\\\hline
$Pt^{5+6}$                &     16.6&     13.1&     11.2&      9.0&      6.6\\\hline
$\Ptgp$&     16.5&     13.1&     11.2&      9.0&      6.7\\\hline
$\Ptgj$ &     15.8&     12.6&     10.7&      8.5&      6.4\\\hline
$\Ptgje$   &     15.7&     12.5&     10.6&      8.4&      6.3\\\hline
$\gJ$                     &   0.0433&   0.0254&   0.0173&   0.0103&   0.0061\\\hline
$1-cos(\Delta \phi)$      &   0.0059&   0.0043&   0.0032&   0.0020&   0.0011\\\hline
$\Dpj$                    &   0.0374&   0.0212&   0.0141&   0.0083&   0.0050\\\hline
                   Entries&    59683&    44691&    34139&    20072&     4019\\\hline
\hline
\end{tabular}

\end{center}
\end{table}

\begin{table}
\begin{center}
\vskip-1cm
\caption{Selection 2. $ \phigj=180^\circ \pm  15^\circ$. UA1 algorithm. }
\vskip0.2cm
\begin{tabular}{||c||c|c|c|c|c|c|c|c|c|c||}
\hline
\hline
$Pt^{clust}_{CUT}$ &$\quad$  30 $\quad$&$\quad$  20 $\quad$&$\quad$ 15 $\quad$&$\quad$  10 $\quad$&$\quad\ $   5 $\quad\ $\\\hline
\hline
Nevent                    &    37984&    31376&    25953&    17917&     4672\\\hline
$Pt56$                    &     17.7&     14.9&     13.2&     11.1&      8.6\\\hline
$\Delta \phi$             &      4.6&      3.9&      3.5&      2.9&      2.1\\\hline
$Pt^{out}$                &     13.7&     11.5&     10.1&      8.3&      6.0\\\hline
$Pt^{\eta > 5}$           &      4.7&      4.7&      4.7&      4.7&      4.5\\\hline
$Pt^{5+6}$                &     14.0&     11.7&     10.3&      8.7&      6.7\\\hline
$\Ptgp$&     14.3&     12.0&     10.5&      8.8&      6.9\\\hline
$\Ptgj$ &     13.8&     11.6&     10.2&      8.5&      6.7\\\hline
$\Ptgje$   &     13.8&     11.7&     10.2&      8.5&      6.7\\\hline
$\gJ$                     &  -0.0057&  -0.0060&  -0.0052&  -0.0036&   0.0006\\\hline
$1-cos(\Delta \phi)$      &   0.0053&   0.0040&   0.0032&   0.0022&   0.0014\\\hline
$\Dpj$                    &  -0.0110&  -0.0100&  -0.0083&  -0.0057&  -0.0007\\\hline
                   Entries&    19803&    16358&    13531&     9341&     2436\\\hline
\hline
\end{tabular}
\vskip0.8cm
\caption{Selection 2. $ \phigj=180^\circ \pm  15^\circ$. LUCELL algorithm. }
\vskip0.2cm
\begin{tabular}{||c||c|c|c|c|c|c|c|c|c|c||}
\hline
\hline
$Pt^{clust}_{CUT}$ &$\quad$  30 $\quad$&$\quad$  20 $\quad$&$\quad$ 15 $\quad$&$\quad$  10 $\quad$&$\quad\ $   5 $\quad\ $\\\hline
\hline
Nevent                    &    36338&    29489&    23986&    15690&     4157\\\hline
$Pt56$                    &     17.3&     14.3&     12.5&     10.4&      7.9\\\hline
$\Delta \phi$             &      4.5&      3.8&      3.3&      2.6&      1.9\\\hline
$Pt^{out}$                &     13.5&     11.2&      9.6&      7.7&      5.5\\\hline
$Pt^{\eta > 5}$           &      4.7&      4.7&      4.7&      4.6&      4.4\\\hline
$Pt^{5+6}$                &     13.7&     11.2&      9.7&      8.0&      6.0\\\hline
$\Ptgp$&     13.9&     11.5&      9.9&      8.2&      6.1\\\hline
$\Ptgj$ &     13.6&     11.3&      9.7&      7.9&      6.0\\\hline
$\Ptgje$   &     13.6&     11.3&      9.7&      7.9&      6.0\\\hline
$\gJ$                     &  -0.0061&  -0.0063&  -0.0054&  -0.0043&  -0.0013\\\hline
$1-cos(\Delta \phi)$      &   0.0052&   0.0038&   0.0029&   0.0019&   0.0011\\\hline
$\Dpj$                    &  -0.0113&  -0.0101&  -0.0083&  -0.0061&  -0.0023\\\hline
                   Entries&    18945&    15374&    12505&     8180&     2168\\\hline
\hline
\end{tabular}
\end{center}
\end{table}

\begin{table}
\begin{center}
\vskip-1cm
\caption{Selection 3. $ \phigj=180^\circ \pm  15^\circ$. UA1 algorithm. }
\vskip0.2cm
\begin{tabular}{||c||c|c|c|c|c|c|c|c|c|c||}
\hline
\hline
$Pt^{clust}_{CUT}$ &$\quad$  30 $\quad$&$\quad$  20 $\quad$&$\quad$ 15 $\quad$&$\quad$  10 $\quad$&$\quad\ $   5 $\quad\ $\\\hline
\hline
Nevent                    &    34158&    27425&    22067&    13807&     3242\\\hline
$Pt56$                    &     15.7&     12.8&     11.0&      9.1&      7.2\\\hline
$\Delta \phi$             &      4.2&      3.4&      2.9&      2.3&      1.8\\\hline
$Pt^{out}$                &     12.3&      9.8&      8.4&      6.6&      4.6\\\hline
$Pt^{\eta > 5}$           &      4.7&      4.7&      4.7&      4.6&      4.6\\\hline
$Pt^{5+6}$                &     12.3&     10.0&      8.5&      6.9&      5.4\\\hline
$\Ptgp$&     12.6&     10.2&      8.8&      7.1&      5.5\\\hline
$\Ptgj$ &     12.4&     10.0&      8.6&      7.0&      5.9\\\hline
$\Ptgje$   &     12.4&     10.1&      8.6&      7.0&      5.9\\\hline
$\gJ$                     &  -0.0060&  -0.0048&  -0.0043&  -0.0017&   0.0034\\\hline
$1-cos(\Delta \phi)$      &   0.0045&   0.0031&   0.0023&   0.0015&   0.0011\\\hline
$\Dpj$                    &  -0.0105&  -0.0079&  -0.0066&  -0.0032&   0.0023\\\hline
                   Entries&    17808&    14298&     11505&    7198&    1691\\\hline
\hline
\end{tabular}
\vskip0.8cm
\caption{Selection 3. $ \phigj=180^\circ \pm  15^\circ$. LUCELL algorithm. }
\vskip0.2cm
\begin{tabular}{||c||c|c|c|c|c|c|c|c|c|c||}
\hline
\hline
$Pt^{clust}_{CUT}$ &$\quad$  30 $\quad$&$\quad$  20 $\quad$&$\quad$ 15 $\quad$&$\quad$  10 $\quad$&$\quad\ $   5 $\quad\ $\\\hline
\hline
Nevent                    &    34158&    27425&    22067&    13807&     3242\\\hline
$Pt56$                    &     15.7&     12.8&     11.0&      9.1&      7.2\\\hline
$\Delta \phi$             &      4.2&      3.4&      2.9&      2.3&      1.8\\\hline
$Pt^{out}$                &     12.3&      9.9&      8.4&      6.7&      4.6\\\hline
$Pt^{\eta > 5}$           &      4.7&      4.7&      4.7&      4.6&      4.6\\\hline
$Pt^{5+6}$                &     12.3&     10.0&      8.5&      6.9&      5.4\\\hline
$\Ptgp$&     12.6&     10.2&      8.8&      7.1&      5.5\\\hline
$\Ptgj$ &     12.4&     10.1&      8.6&      7.0&      5.9\\\hline
$\Ptgje$   &     12.5&     10.1&      8.7&      7.0&      5.9\\\hline
$\gJ$                     &  -0.0062&  -0.0053&  -0.0049&  -0.0024&   0.0023\\\hline
$1-cos(\Delta \phi)$      &   0.0045&   0.0031&   0.0023&   0.0015&   0.0011\\\hline
$\Dpj$                    &  -0.0107&  -0.0084&  -0.0071&  -0.0038&   0.0013\\\hline
                   Entries&    17808&    14298&     11505&    7198&    1691\\\hline
\hline
\end{tabular}
\end{center}
\end{table}

\setcounter{table}{0}


\begin{table}
\large{\bf Appendix 3}\\
\begin{center}
\large{ $200 < \Pt^{\gamma} < 240 ~GeV/c$}\\[20pt]
\normalsize
\caption{Selection 1. $ \phigj=180^\circ \pm  180^\circ$. UA1 algorithm. }
\vskip0.2cm
\begin{tabular}{||c||c|c|c|c|c|c|c|c|c|c||}
\hline
\hline
$Pt^{clust}_{CUT}$ &$\quad$  30 $\quad$&$\quad$  20 $\quad$&$\quad$ 15 $\quad$&$\quad$  10 $\quad$&$\quad\ $   5 $\quad\ $\\\hline
\hline
Nevent                    &     9389&     6714&     5065&     3057&      559\\\hline
$Pt56$                    &     24.5&     19.7&     17.2&     14.3&     10.8\\\hline
$\Delta \phi$             &      3.2&      2.4&      2.0&      1.6&      1.2\\\hline
$Pt^{out}$                &     18.5&     14.1&     11.7&      9.3&      6.8\\\hline
$Pt^{\eta > 5}$           &      4.8&      4.8&      4.7&      4.7&      4.6\\\hline
$Pt^{5+6}$                &     20.0&     15.9&     13.8&     11.4&      8.7\\\hline
$\Ptgp$&     20.2&     16.1&     14.0&     11.6&      8.8\\\hline
$\Ptgj$ &     19.1&     14.8&     12.5&     10.1&      7.8\\\hline
$\Ptgje$   &     19.0&     14.7&     12.4&     10.0&      7.8\\\hline
$\gJ$                     &   0.0202&   0.0131&   0.0097&   0.0066&   0.0042\\\hline
$1-cos(\Delta \phi)$      &   0.0029&   0.0016&   0.0011&   0.0007&   0.0005\\\hline
$\Dpj$                    &   0.0173&   0.0115&   0.0086&   0.0059&   0.0037\\\hline
                   Entries&    52803&    37761&    28486&    17192&     3143\\\hline
\hline
\end{tabular}
\vskip0.8cm
\caption{Selection 1. $ \phigj=180^\circ \pm  180^\circ$. LUCELL algorithm. }
\vskip0.2cm
\begin{tabular}{||c||c|c|c|c|c|c|c|c|c|c||}
\hline
\hline
$Pt^{clust}_{CUT}$ &$\quad$  30 $\quad$&$\quad$  20 $\quad$&$\quad$ 15 $\quad$&$\quad$  10 $\quad$&$\quad\ $   5 $\quad\ $\\\hline
\hline
Nevent                    &     8788&     6185&     4549&     2589&      495\\\hline
$Pt56$                    &     23.9&     18.9&     16.2&     13.1&      9.3\\\hline
$\Delta \phi$             &      3.1&      2.3&      1.9&      1.5&      1.1\\\hline
$Pt^{out}$                &     17.8&     13.3&     10.9&      8.4&      5.8\\\hline
$Pt^{\eta > 5}$           &      4.8&      4.8&      4.7&      4.7&      4.5\\\hline
$Pt^{5+6}$                &     19.5&     15.2&     12.9&     10.3&      7.2\\\hline
$\Ptgp$&     19.6&     15.3&     13.0&     10.5&      7.4\\\hline
$\Ptgj$ &     18.4&     14.0&     11.6&      9.1&      6.7\\\hline
$\Ptgje$   &     18.3&     13.9&     11.5&      9.1&      6.7\\\hline
$\gJ$                     &   0.0174&   0.0101&   0.0069&   0.0039&   0.0023\\\hline
$1-cos(\Delta \phi)$      &   0.0026&   0.0014&   0.0009&   0.0006&   0.0003\\\hline
$\Dpj$                    &   0.0148&   0.0087&   0.0060&   0.0034&   0.0021\\\hline
                   Entries&    49425&    34786&    25582&    14562&     2786\\\hline
\hline
\end{tabular}
\end{center}
\end{table}

\begin{table}
\begin{center}
\vskip-1cm
\caption{Selection 1. $ \phigj=180^\circ \pm  15^\circ$. UA1 algorithm. }
\vskip0.2cm
\begin{tabular}{||c||c|c|c|c|c|c|c|c|c|c||}
\hline
\hline
$Pt^{clust}_{CUT}$ &$\quad$  30 $\quad$&$\quad$  20 $\quad$&$\quad$ 15 $\quad$&$\quad$  10 $\quad$&$\quad\ $   5 $\quad\ $\\\hline
\hline
Nevent                    &     9343&     6711&     5064&     3056&      559\\\hline
$Pt56$                    &     24.3&     19.7&     17.1&     14.3&     10.8\\\hline
$\Delta \phi$             &      3.1&      2.4&      2.0&      1.6&      1.2\\\hline
$Pt^{out}$                &     18.2&     14.0&     11.7&      9.3&      6.8\\\hline
$Pt^{\eta > 5}$           &      4.8&      4.8&      4.7&      4.7&      4.6\\\hline
$Pt^{5+6}$                &     19.8&     15.9&     13.7&     11.4&      8.7\\\hline
$\Ptgp$&     20.0&     16.0&     13.9&     11.6&      8.8\\\hline
$\Ptgj$ &     18.9&     14.7&     12.4&     10.1&      7.8\\\hline
$\Ptgje$   &     18.7&     14.6&     12.4&     10.0&      7.7\\\hline
$\gJ$                     &   0.0198&   0.0130&   0.0097&   0.0066&   0.0041\\\hline
$1-cos(\Delta \phi)$      &   0.0026&   0.0016&   0.0011&   0.0007&   0.0005\\\hline
$\Dpj$                    &   0.0172&   0.0115&   0.0086&   0.0059&   0.0037\\\hline
                   Entries&    52542&    37741&    28477&    17189&     3142\\\hline
\hline
\end{tabular}
\vskip0.8cm
\caption{Selection 1. $ \phigj=180^\circ \pm  15^\circ$. LUCELL algorithm. }
\vskip0.2cm
\begin{tabular}{||c||c|c|c|c|c|c|c|c|c|c||}
\hline
\hline
$Pt^{clust}_{CUT}$ &$\quad$  30 $\quad$&$\quad$  20 $\quad$&$\quad$ 15 $\quad$&$\quad$  10 $\quad$&$\quad\ $   5 $\quad\ $\\\hline
\hline
Nevent                    &     8758&     6183&     4549&     2589&      503\\\hline
$Pt56$                    &     23.8&     18.9&     16.2&     13.1&      9.3\\\hline
$\Delta \phi$             &      3.0&      2.3&      1.9&      1.5&      1.1\\\hline
$Pt^{out}$                &     17.6&     13.3&     10.9&      8.4&      5.8\\\hline
$Pt^{\eta > 5}$           &      4.8&      4.7&      4.7&      4.7&      4.5\\\hline
$Pt^{5+6}$                &     19.3&     15.1&     12.9&     10.3&      7.2\\\hline
$\Ptgp$&     19.5&     15.3&     13.0&     10.5&      7.4\\\hline
$\Ptgj$ &     18.2&     14.0&     11.6&      9.1&      6.7\\\hline
$\Ptgje$   &     18.1&     13.9&     11.5&      9.1&      6.7\\\hline
$\gJ$                     &   0.0172&   0.0101&   0.0069&   0.0039&   0.0023\\\hline
$1-cos(\Delta \phi)$      &   0.0025&   0.0014&   0.0009&   0.0006&   0.0003\\\hline
$\Dpj$                    &   0.0147&   0.0087&   0.0060&   0.0034&   0.0021\\\hline
                   Entries&    49253&    34775&    25582&    14562&     2786\\\hline
\hline
\end{tabular}

\end{center}
\end{table}

\begin{table}
\begin{center}
\vskip-1cm
\caption{Selection 2. $ \phigj=180^\circ \pm  15^\circ$. UA1 algorithm. }
\vskip0.2cm
\begin{tabular}{||c||c|c|c|c|c|c|c|c|c|c||}
\hline
\hline
$Pt^{clust}_{CUT}$ &$\quad$  30 $\quad$&$\quad$  20 $\quad$&$\quad$ 15 $\quad$&$\quad$  10 $\quad$&$\quad\ $   5 $\quad\ $\\\hline
\hline
Nevent                    &     5797&     4466&     3558&     2340&      509\\\hline
$Pt56$                    &     21.6&     17.7&     15.5&     13.2&     10.4\\\hline
$\Delta \phi$             &      2.9&      2.3&      1.9&      1.6&      1.2\\\hline
$Pt^{out}$                &     16.3&     12.9&     10.9&      8.9&      6.7\\\hline
$Pt^{\eta > 5}$           &      4.8&      4.8&      4.7&      4.7&      4.6\\\hline
$Pt^{5+6}$                &     17.2&     14.0&     12.2&     10.4&      8.3\\\hline
$\Ptgp$&     17.6&     14.3&     12.5&     10.7&      8.5\\\hline
$\Ptgj$ &     16.8&     13.4&     11.5&      9.5&      7.6\\\hline
$\Ptgje$   &     16.8&     13.4&     11.5&      9.5&      7.6\\\hline
$\gJ$                     &   0.0014&   0.0010&   0.0015&   0.0019&   0.0030\\\hline
$1-cos(\Delta \phi)$      &   0.0023&   0.0014&   0.0010&   0.0007&   0.0005\\\hline
$\Dpj$                    &  -0.0009&  -0.0004&   0.0005&   0.0012&   0.0025\\\hline
                   Entries&    32603&    25116&    20009&    13160&     2862\\\hline
\hline
\end{tabular}
\vskip0.8cm
\caption{Selection 2. $ \phigj=180^\circ \pm  15^\circ$. LUCELL algorithm. }
\vskip0.2cm
\begin{tabular}{||c||c|c|c|c|c|c|c|c|c|c||}
\hline
\hline
$Pt^{clust}_{CUT}$ &$\quad$  30 $\quad$&$\quad$  20 $\quad$&$\quad$ 15 $\quad$&$\quad$  10 $\quad$&$\quad\ $   5 $\quad\ $\\\hline
\hline
Nevent                    &     5433&     4114&     3190&     1971&      442\\\hline
$Pt56$                    &     21.0&     17.0&     14.6&     12.0&      8.7\\\hline
$\Delta \phi$             &      2.8&      2.2&      1.8&      1.4&      1.1\\\hline
$Pt^{out}$                &     15.9&     12.4&     10.3&      8.1&      5.7\\\hline
$Pt^{\eta > 5}$           &      4.8&      4.7&      4.7&      4.7&      4.5\\\hline
$Pt^{5+6}$                &     16.7&     13.4&     11.5&      9.3&      6.7\\\hline
$\Ptgp$&     17.1&     13.7&     11.8&      9.6&      6.8\\\hline
$\Ptgj$ &     16.4&     12.8&     10.8&      8.7&      6.6\\\hline
$\Ptgje$   &     16.4&     12.9&     10.8&      8.7&      6.6\\\hline
$\gJ$                     &   0.0005&  -0.0002&   0.0001&   0.0000&   0.0009\\\hline
$1-cos(\Delta \phi)$      &   0.0022&   0.0013&   0.0009&   0.0005&   0.0003\\\hline
$\Dpj$                    &  -0.0016&  -0.0015&  -0.0007&  -0.0005&   0.0006\\\hline
                   Entries&    30556&    23138&    17939&    11085&     2447\\\hline
\hline
\end{tabular}
\end{center}
\end{table}

\begin{table}
\begin{center}
\vskip-1cm
\caption{Selection 3. $ \phigj=180^\circ \pm  15^\circ$. UA1 algorithm. }
\vskip0.2cm
\begin{tabular}{||c||c|c|c|c|c|c|c|c|c|c||}
\hline
\hline
$Pt^{clust}_{CUT}$ &$\quad$  30 $\quad$&$\quad$  20 $\quad$&$\quad$ 15 $\quad$&$\quad$  10 $\quad$&$\quad\ $   5 $\quad\ $\\\hline
\hline
Nevent                    &     5053&     3826&     2935&     2735&      327\\\hline
$Pt56$                    &     18.2&     14.6&     12.7&     10.6&      7.9\\\hline
$\Delta \phi$             &      2.4&      1.9&      1.6&      1.3&      1.0\\\hline
$Pt^{out}$                &     13.6&     10.5&      8.8&      7.2&      4.8\\\hline
$Pt^{\eta > 5}$           &      4.7&      4.7&      4.7&      4.7&      4.6\\\hline
$Pt^{5+6}$                &     14.4&     11.3&      9.8&      8.2&      6.1\\\hline
$\Ptgp$&     14.7&     11.6&     10.1&      8.4&      6.2\\\hline
$\Ptgj$ &     14.2&     11.1&      9.4&      8.0&      6.3\\\hline
$\Ptgje$   &     14.2&     11.1&      9.4&      8.0&      6.2\\\hline
$\gJ$                     &   0.0002&   0.0004&   0.0012&   0.0013&   0.0024\\\hline
$1-cos(\Delta \phi)$      &   0.0016&   0.0009&   0.0006&   0.0005&   0.0003\\\hline
$\Dpj$                    &  -0.0014&  -0.0005&   0.0006&   0.0009&   0.0021\\\hline
                   Entries&    28417&    21518&   16504&    9755&     1811\\\hline
\hline
\end{tabular}
\vskip0.8cm
\caption{Selection 3. $ \phigj=180^\circ \pm  15^\circ$. LUCELL algorithm. }
\vskip0.2cm
\begin{tabular}{||c||c|c|c|c|c|c|c|c|c|c||}
\hline
\hline
$Pt^{clust}_{CUT}$ &$\quad$  30 $\quad$&$\quad$  20 $\quad$&$\quad$ 15 $\quad$&$\quad$  10 $\quad$&$\quad\ $   5 $\quad\ $\\\hline
\hline
Nevent                    &     5053&     3826&     2935&     2735&      327\\\hline
$Pt56$                    &     18.2&     14.6&     12.7&     10.6&      7.9\\\hline
$\Delta \phi$             &      2.4&      1.9&      1.6&      1.3&      1.0\\\hline
$Pt^{out}$                &     13.7&     10.5&      8.8&      7.1&      4.7\\\hline
$Pt^{\eta > 5}$           &      4.7&      4.7&      4.7&      4.7&      4.6\\\hline
$Pt^{5+6}$                &     14.4&     11.3&      9.8&      8.2&      6.1\\\hline
$\Ptgp$&     14.7&     11.6&     10.1&      8.4&      6.2\\\hline
$\Ptgj$ &     14.3&     11.1&      9.4&      8.0&      6.2\\\hline
$\Ptgje$   &     14.3&     11.1&      9.5&      8.0&      6.2\\\hline
$\gJ$                     &  -0.0005&  -0.0004&   0.0002&   0.0000&   0.0011\\\hline
$1-cos(\Delta \phi)$      &   0.0016&   0.0009&   0.0006&   0.0005&   0.0003\\\hline
$\Dpj$                    &  -0.0021&  -0.0013&  -0.0005&  -0.0005&   0.0008\\\hline
                  Entries&    28417&    21518&   16504&    9755&     1811\\\hline
\hline
\end{tabular}

\end{center}
\end{table}

\setcounter{table}{0}
\begin{table}
\large{\bf Appendix 4}\\
\begin{center}
\large{ $300 < \Pt^{\gamma} < 360 ~GeV/c$}\\[20pt]
\normalsize
\caption{Selection 1. $ \phigj=180^\circ \pm  180^\circ$. UA1 algorithm. }
\vskip0.2cm
\begin{tabular}{||c||c|c|c|c|c|c|c|c|c|c||}
\hline
\hline
$Pt^{clust}_{CUT}$ &$\quad$  30 $\quad$&$\quad$  20 $\quad$&$\quad$ 15 $\quad$&$\quad$  10 $\quad$&$\quad\ $   5 $\quad\ $\\\hline
\hline
Nevent                    &     1872&     1315&      977&      579&      107\\\hline
$Pt56$                    &     26.7&     21.7&     18.8&     15.7&     12.7\\\hline
$\Delta \phi$             &      2.2&      1.7&      1.4&      1.1&      0.8\\\hline
$Pt^{out}$                &     19.0&     14.5&     12.1&      9.7&      7.0\\\hline
$Pt^{\eta > 5}$           &      4.8&      4.7&      4.7&      4.7&      4.6\\\hline
$Pt^{5+6}$                &     21.8&     17.6&     15.2&     12.7&     10.4\\\hline
$\Ptgp$&     22.1&     17.9&     15.5&     13.0&     10.6\\\hline
$\Ptgj$ &     19.9&     15.5&     13.1&     10.7&      8.6\\\hline
$\Ptgje$   &     19.8&     15.4&     13.0&     10.6&      8.6\\\hline
$\gJ$                     &   0.0124&   0.0083&   0.0059&   0.0042&   0.0029\\\hline
$1-cos(\Delta \phi)$      &   0.0013&   0.0008&   0.0005&   0.0003&   0.0002\\\hline
$\Dpj$                    &   0.0111&   0.0076&   0.0054&   0.0039&   0.0027\\\hline
                   Entries&    46306&    32515&    24159&    14319&     2642\\\hline
\hline
\end{tabular}
\vskip0.8cm
\caption{Selection 1. $ \phigj=180^\circ \pm  180^\circ$. LUCELL algorithm. }
\vskip0.2cm
\begin{tabular}{||c||c|c|c|c|c|c|c|c|c|c||}
\hline
\hline
$Pt^{clust}_{CUT}$ &$\quad$  30 $\quad$&$\quad$  20 $\quad$&$\quad$ 15 $\quad$&$\quad$  10 $\quad$&$\quad\ $   5 $\quad\ $\\\hline
\hline
Nevent                    &     1752&     1204&      878&      489&       94 \\\hline
$Pt56$                    &     26.0&     20.7&     17.7&     14.2&     10.9\\\hline
$\Delta \phi$             &      2.1&      1.6&      1.3&      1.0&      0.7\\\hline
$Pt^{out}$                &     18.4&     13.7&     11.3&      8.6&      5.9\\\hline
$Pt^{\eta > 5}$           &      4.8&      4.8&      4.7&      4.6&      4.5\\\hline
$Pt^{5+6}$                &     21.2&     16.7&     14.2&     11.3&      8.7\\\hline
$\Ptgp$&     21.5&     17.0&     14.4&     11.5&      8.9\\\hline
$\Ptgj$ &     19.3&     14.7&     12.3&      9.7&      7.4\\\hline
$\Ptgje$   &     19.1&     14.6&     12.2&      9.6&      7.4\\\hline
$\gJ$                     &   0.0107&   0.0066&   0.0046&   0.0027&   0.0016\\\hline
$1-cos(\Delta \phi)$      &   0.0012&   0.0007&   0.0004&   0.0003&   0.0001\\\hline
$\Dpj$                    &   0.0095&   0.0060&   0.0041&   0.0025&   0.0015\\\hline
                   Entries&    43323&    29783&    21707&    12104&     2334\\\hline
\hline
\end{tabular}
\end{center}
\end{table}

\begin{table}
\begin{center}
\vskip-1cm
\caption{Selection 1. $ \phigj=180^\circ \pm  15^\circ$. UA1 algorithm. }
\vskip0.2cm
\begin{tabular}{||c||c|c|c|c|c|c|c|c|c|c||}
\hline
\hline
$Pt^{clust}_{CUT}$ &$\quad$  30 $\quad$&$\quad$  20 $\quad$&$\quad$ 15 $\quad$&$\quad$  10 $\quad$&$\quad\ $   5 $\quad\ $\\\hline
\hline
Nevent                    &     1872&     1315&      977&      579&      107\\\hline
$Pt56$                    &     26.7&     21.7&     18.8&     15.7&     12.7\\\hline
$\Delta \phi$             &      2.2&      1.7&      1.4&      1.1&      0.8\\\hline
$Pt^{out}$                &     19.0&     14.5&     12.1&      9.7&      7.0\\\hline
$Pt^{\eta > 5}$           &      4.8&      4.7&      4.7&      4.7&      4.6\\\hline
$Pt^{5+6}$                &     21.8&     17.6&     15.2&     12.7&     10.4\\\hline
$\Ptgp$&     22.1&     17.9&     15.5&     12.9&     10.6\\\hline
$\Ptgj$ &     19.9&     15.5&     13.1&     10.7&      8.6\\\hline
$\Ptgje$   &     19.7&     15.4&     13.0&     10.6&      8.6\\\hline
$\gJ$                     &   0.0124&   0.0083&   0.0059&   0.0042&   0.0029\\\hline
$1-cos(\Delta \phi)$      &   0.0013&   0.0008&   0.0005&   0.0003&   0.0002\\\hline
$\Dpj$                    &   0.0111&   0.0076&   0.0054&   0.0039&   0.0027\\\hline
                   Entries&    46297&    32513&    24157&    14318&     2642\\\hline
\hline
\end{tabular}
\vskip0.8cm
\caption{Selection 1. $ \phigj=180^\circ \pm  15^\circ$. LUCELL algorithm. }
\vskip0.2cm
\begin{tabular}{||c||c|c|c|c|c|c|c|c|c|c||}
\hline
\hline
$Pt^{clust}_{CUT}$ &$\quad$  30 $\quad$&$\quad$  20 $\quad$&$\quad$ 15 $\quad$&$\quad$  10 $\quad$&$\quad\ $   5 $\quad\ $\\\hline
\hline
Nevent                    &     1752&     1204&      878&      489&       96\\\hline
$Pt56$                    &     26.0&     20.7&     17.7&     14.2&     10.9\\\hline
$\Delta \phi$             &      2.1&      1.6&      1.3&      1.0&      0.7\\\hline
$Pt^{out}$                &     18.4&     13.7&     11.3&      8.6&      5.9\\\hline
$Pt^{\eta > 5}$           &      4.8&      4.8&      4.7&      4.6&      4.5\\\hline
$Pt^{5+6}$                &     21.1&     16.7&     14.2&     11.3&      8.7\\\hline
$\Ptgp$&     21.5&     17.0&     14.4&     11.5&      8.9\\\hline
$\Ptgj$ &     19.2&     14.7&     12.3&      9.7&      7.4\\\hline
$\Ptgje$   &     19.1&     14.6&     12.2&      9.6&      7.4\\\hline
$\gJ$                     &   0.0107&   0.0066&   0.0046&   0.0027&   0.0016\\\hline
$1-cos(\Delta \phi)$      &   0.0012&   0.0007&   0.0004&   0.0003&   0.0001\\\hline
$\Dpj$                    &   0.0095&   0.0060&   0.0041&   0.0025&   0.0015\\\hline
                   Entries&    43320&    29783&    21707&    12104&     2334\\\hline
\hline
\end{tabular}

\end{center}
\end{table}

\begin{table}
\begin{center}
\vskip-1cm
\caption{Selection 2. $ \phigj=180^\circ \pm  15^\circ$. UA1 algorithm. }
\vskip0.2cm
\begin{tabular}{||c||c|c|c|c|c|c|c|c|c|c||}
\hline
\hline
$Pt^{clust}_{CUT}$ &$\quad$  30 $\quad$&$\quad$  20 $\quad$&$\quad$ 15 $\quad$&$\quad$  10 $\quad$&$\quad\ $   5 $\quad\ $\\\hline
\hline
Nevent                    &     1464&     1096&      854&      538&      106\\\hline
$Pt56$                    &     24.5&     20.3&     17.9&     15.2&     12.5\\\hline
$\Delta \phi$             &      2.1&      1.6&      1.4&      1.1&      0.8\\\hline
$Pt^{out}$                &     17.4&     13.7&     11.6&      9.5&      7.0\\\hline
$Pt^{\eta > 5}$           &      4.7&      4.7&      4.7&      4.7&      4.6\\\hline
$Pt^{5+6}$                &     19.7&     16.3&     14.3&     12.2&     10.2\\\hline
$\Ptgp$&     20.2&     16.7&     14.7&     12.5&     10.5\\\hline
$\Ptgj$ &     18.2&     14.5&     12.6&     10.5&      8.6\\\hline
$\Ptgje$   &     18.2&     14.5&     12.5&     10.4&      8.6\\\hline
$\gJ$                     &   0.0034&   0.0032&   0.0028&   0.0029&   0.0028\\\hline
$1-cos(\Delta \phi)$      &   0.0012&   0.0007&   0.0005&   0.0003&   0.0002\\\hline
$\Dpj$                    &   0.0022&   0.0025&   0.0023&   0.0026&   0.0026\\\hline
                   Entries&    36198&    27104&    21115&    13303&     2610\\\hline
\hline
\end{tabular}
\vskip0.8cm
\caption{Selection 2. $ \phigj=180^\circ \pm  15^\circ$. LUCELL algorithm. }
\vskip0.2cm
\begin{tabular}{||c||c|c|c|c|c|c|c|c|c|c||}
\hline
\hline
$Pt^{clust}_{CUT}$ &$\quad$  30 $\quad$&$\quad$  20 $\quad$&$\quad$ 15 $\quad$&$\quad$  10 $\quad$&$\quad\ $   5 $\quad\ $\\\hline
\hline
Nevent                    &     1359&      993&      756&      448&       94\\\hline
$Pt56$                    &     23.6&     19.2&     16.6&     13.6&     10.8\\\hline
$\Delta \phi$             &      2.0&      1.5&      1.3&      1.0&      0.7\\\hline
$Pt^{out}$                &     16.9&     13.0&     10.8&      8.4&      5.9\\\hline
$Pt^{\eta > 5}$           &      4.7&      4.7&      4.7&      4.6&      4.5\\\hline
$Pt^{5+6}$                &     19.0&     15.3&     13.2&     10.7&      8.6\\\hline
$\Ptgp$&     19.4&     15.7&     13.5&     11.0&      8.8\\\hline
$\Ptgj$ &     17.8&     13.9&     11.8&      9.4&      7.3\\\hline
$\Ptgje$   &     17.7&     13.9&     11.8&      9.4&      7.3\\\hline
$\gJ$                     &   0.0024&   0.0020&   0.0017&   0.0014&   0.0014\\\hline
$1-cos(\Delta \phi)$      &   0.0011&   0.0006&   0.0004&   0.0003&   0.0001\\\hline
$\Dpj$                    &   0.0013&   0.0014&   0.0013&   0.0011&   0.0013\\\hline
                   Entries&    33599&    24547&    18686&    11070&     2287\\\hline
\hline
\end{tabular}
\end{center}
\end{table}

\begin{table}
\begin{center}
\vskip-1cm
\caption{Selection 3. $ \phigj=180^\circ \pm  15^\circ$. UA1 algorithm. }
\vskip0.2cm
\begin{tabular}{||c||c|c|c|c|c|c|c|c|c|c||}
\hline
\hline
$Pt^{clust}_{CUT}$ &$\quad$  30 $\quad$&$\quad$  20 $\quad$&$\quad$ 15 $\quad$&$\quad$  10 $\quad$&$\quad\ $   5 $\quad\ $\\\hline\hline
Nevent                    &     1264&      925&      696&      394&       70\\\hline
$Pt56$                    &     20.0&     16.3&     14.2&     11.7&      8.8\\\hline
$\Delta \phi$             &      1.7&      1.3&      1.1&      0.9&      0.7\\\hline
$Pt^{out}$                &     14.4&     11.0&      9.2&      7.3&      5.2\\\hline
$Pt^{\eta > 5}$           &      4.7&      4.7&      4.7&      4.6&      4.7\\\hline
$Pt^{5+6}$                &     15.8&     12.9&     11.2&      9.2&      6.7\\\hline
$\Ptgp$&     16.3&     13.2&     11.5&      9.4&      6.9\\\hline
$\Ptgj$ &     15.3&     12.0&     10.4&      8.5&      7.4\\\hline
$\Ptgje$   &     15.3&     12.0&     10.4&      8.5&      7.3\\\hline
$\gJ$                     &   0.0010&   0.0016&   0.0021&   0.0026&   0.00323\\\hline
$1-cos(\Delta \phi)$      &   0.0008&   0.0005&   0.0003&   0.0002&   0.0001\\\hline
$\Dpj$                    &   0.0002&   0.0011&   0.0018&   0.0025&   0.00312\\\hline
                  Entries&    31247&    22829&     17991&    10301&      1692\\\hline
\hline
\end{tabular}
\vskip0.8cm
\caption{Selection 3. $ \phigj=180^\circ \pm  15^\circ$. LUCELL algorithm. }
\vskip0.2cm
\begin{tabular}{||c||c|c|c|c|c|c|c|c|c|c||}
\hline
\hline
$Pt^{clust}_{CUT}$ &$\quad$  30 $\quad$&$\quad$  20 $\quad$&$\quad$ 15 $\quad$&$\quad$  10 $\quad$&$\quad\ $   5 $\quad\ $\\\hline
\hline
Nevent                    &     1264&      925&      696&      394&       70\\\hline
$Pt56$                    &     20.0&     16.3&     14.2&     11.7&      8.8\\\hline
$\Delta \phi$             &      1.7&      1.3&      1.1&      0.9&      0.7\\\hline
$Pt^{out}$                &     14.5&     11.0&      9.2&      7.2&      5.1\\\hline
$Pt^{\eta > 5}$           &      4.7&      4.7&      4.7&      4.6&      4.7\\\hline
$Pt^{5+6}$                &     15.8&     12.9&     11.2&      9.2&      6.7\\\hline
$\Ptgp$&     16.3&     13.2&     11.5&      9.4&      6.9\\\hline
$\Ptgj$ &     15.4&     12.0&     10.3&      8.5&      7.3\\\hline
$\Ptgje$   &     15.4&     12.0&     10.3&      8.4&      7.2\\\hline
$\gJ$                     &   0.0005&   0.0008&   0.0013&   0.0016&   0.0023\\\hline
$1-cos(\Delta \phi)$      &   0.0008&   0.0005&   0.0003&   0.0002&   0.0001\\\hline
$\Dpj$                    &  -0.0003&   0.0004&   0.0010&   0.0014&   0.0022\\\hline
                  Entries&    31247&    22829&     17991&    10301&      1692\\\hline
\hline
\end{tabular}

\end{center}
\end{table}


\begin{thebibliography}{99}
\bibitem{1}
D.V.~Bandourin, V.F.~Konoplyanikov, N.B.~Skachkov.
``Jet energy scale setting with \gpj events at LHC
energies. Generalities, selection rules'', 
JINR Preprint E2-2000-251, JINR, Dubna.
\bibitem{2}
D.V.~Bandourin, V.F.~Konoplyanikov, N.B.~Skachkov.
``Jet energy scale setting with \gpj events at LHC
energies. Event rates, $\Pt$ structure of jet'',
JINR Preprint E2-2000-252, JINR, Dubna.
\bibitem{3}
D.V.~Bandourin, V.F.~Konoplyanikov, N.B.~Skachkov.
``Jet energy scale setting with \gpj events at LHC
energies. Selection of events with a clean \gpj topology and
$\Pt^{\gamma}\! -\! \Pt^{Jet}$ disbalance'',
JINR Preprint E2-2000-254, JINR, Dubna.
\bibitem{4}
D.V.~Bandourin, V.F.~Konoplyanikov, N.B.~Skachkov.
``Jet energy scale setting with \gpj events at LHC
energies. Detailed study of the background suppression''.
JINR Preprint E2-2000-255, JINR, Dubna.

\end{thebibliography}
\end{document}